\documentclass[conference]{IEEEtran}
\IEEEoverridecommandlockouts
\usepackage{cite}
\usepackage{amsmath,amssymb,amsfonts}
\usepackage{algorithmic}
\usepackage{graphicx}
\usepackage{textcomp}
\usepackage{xcolor}
 
\usepackage{url}

\usepackage{float}
\usepackage{array}  % 对齐
\usepackage{booktabs} %调整表格线与上下内容的间隔
\usepackage{multirow}
\usepackage{makecell}
\usepackage{bbding} % 用于打对勾
\def\BibTeX{{\rm B\kern-.05em{\sc i\kern-.025em b}\kern-.08em
    T\kern-.1667em\lower.7ex\hbox{E}\kern-.125emX}}
\begin{document}

% Contrastive event-object alignment aided audio-visual scene classification
\title{Audio-visual scene classification via contrastive event-object alignment and semantic-based fusion
}

\author{
\IEEEauthorblockN{Yuanbo Hou}
\IEEEauthorblockA{
\textit{WAVES Research Group} \\
\textit{Ghent University}, 
Gent, Belgium \\
Yuanbo.Hou@UGent.be}
\and
\IEEEauthorblockN{Bo Kang}
\IEEEauthorblockA{
\textit{IDLAB} \\
\textit{Ghent University},
Gent, Belgium \\
Bo.Kang@UGent.be}
\and 
\IEEEauthorblockN{Dick Botteldooren}
\IEEEauthorblockA{
\textit{WAVES Research Group} \\
\textit{Ghent University},
Gent, Belgium \\
Dick.Botteldooren@UGent.be}
}

\maketitle

\begin{abstract}
Previous works on scene classification are mainly based on audio or visual signals, while humans perceive the environmental scenes through multiple senses.
% (auditory and visual). 
% Similar to humans, intelligent robots with the awareness of audio-visual environmental information, can also more accurately identify the scenes they are in.
Recent studies on audio-visual scene classification separately fine-tune the large-scale audio and image pre-trained models on the target dataset, then either fuse the intermediate representations of the audio model and the visual model, or fuse the coarse-grained decision of both models at the clip level.
Such methods ignore the detailed
% rich and fine-grained 
audio events and visual objects in audio-visual scenes (AVS), while humans often identify a scene through both audio events and visual objects within, and the congruence between them.
To exploit the fine-grained information of audio events and visual objects in AVS, and coordinate the implicit relationship between audio events and visual objects, 
this paper proposes a multi-branch model equipped with contrastive event-object alignment (CEOA) and semantic-based fusion (SF) for AVSC.
CEOA aims to align the learned embeddings of audio events and visual objects by comparing the difference between audio-visual event-object pairs.
Then, visual objects associated with certain audio events and vice versa are accentuated by cross-attention and undergo SF for semantic-level fusion.
Experiments show that:
1) the proposed AVSC model equipped with CEOA and SF outperforms the results of audio-only and visual-only models, i.e., the audio-visual results are better than the results from a single modality.
2) CEOA 
% effectively 
aligns the embeddings of audio events and related visual objects on a fine-grained level, and the SF effectively integrates both;
3) Compared with other large-scale integrated systems, the proposed model shows competitive performance, even without using additional datasets and data augmentation tricks.

% the representations are fused
% by attention based on semantic similarity to shape the acous-
% tic representations through the probability of anchor vocaliza-
% tion. 

\end{abstract}


\vspace{-0.5cm}
\section{Introduction}
Audio-visual scene classification (AVSC) aims to use both audio and visual modalities to classify a video recording into one of the predefined scene categories (such as metro station, airport, or street pedestrian).
Compared with scene classification relying solely on audio or visual modality, AVSC is able to not only exploit the richer information from the data but also leverage the relationship between the two modals to achieve better accuracy.
Recently, AVSC has attracted many interests due to its wide applications  \cite{audio-visualevent}\cite{video_description}\cite{robot1}\cite{robot2}.
% such as audio-visual event surveillance \cite{audio-visualevent}, scene-aware dialog \cite{dialog}, video description generation \cite{video_description}, and environment perception \cite{robot1}\cite{robot2}.
% , as well as scene characterization \cite{scene_characterization}.

%%%
Scene classification provides semantic information to effectively guide higher-level audio or visual content understanding. 
% Previous studies
Prior works
\cite{audio_scene1}\cite{audio_scene2}\cite{image_scene}\cite{image_scene2} 
% \cite{audio_scene3}\cite{audio_scene4}\cite{audio_scene5}
% \cite{image_scene}\cite{image_scene2} 
on scene classification are mainly based on either audio or image information, and the related tasks are called acoustic scene classification (ASC) or image scene classification (ISC), respectively.
% The input to an ASC model is usually the acoustic features of audio clips. 
% The model then identifies and distinguishes different scenes based on the high-level representations learned from the acoustic features. 
% In other words, 
The ASC models in these works  \cite{audio_scene1}\cite{audio_scene2}
% \cite{audio_scene3}\cite{audio_scene4}\cite{audio_scene5} 
make decisions based on the clip-level information about scenes. 
In real life, an acoustic scene and the audio events took place within are naturally correlated. For example, in a park scene, birds singing and dogs barking are more likely to occur than keyboard sounds, where the later are often found in the office scenes.
To exploit the inherent relationships between the coarse-grained scenes and corresponding fine-grained events, relation-guided ASC \cite{RGASC} coordinates scene-event relationships for the mutual benefit of scene and event recognition. 
% In comparison, the input of the ISC model 
For ISC models \cite{cheng2018scene}\cite{kim2018hierarchy}, the input is usually an image or image sequence \cite{image_sequence}, and then the scene is recognized based on the rich object information, spatial layout information, as well as the relationship between the objects and layouts.
% In the real world, different scenes could contain common objects, for example, cats seen in a park scene may also appear in a street scene, or in a pet store. 
% Relying only on object information could limit a model's ability to identify complex and diverse scenes thus weaken its discriminative power among scenes.
% To focus on the information of the same objects existing in different scenes, semantic-based image descriptor \cite{cheng2018scene} is presented for scene recognition, which utilizes the co-occurrence pattern between objects and different scenes to enhance the inter-class discriminability among diverse scenes.
% Unlike reducing the inter-class similarity between different scenes, a hierarchy of specialist networks \cite{kim2018hierarchy} is proposed to disentangle the intra-class variation to promote the intra-class generalized performance.
ASC and ISC aim to understand scene semantic information from the perspective of human cognition based on either audio or visual information, while humans often use audio-visual information to distinguish various scenes.
% at the same time.

% Researches in cognitive science \cite{oliva2005gist} reveal that human can perceive scenes in a very short period of time.
% A simple observation is that 
Human naturally recognizes diverse scenes based on various objects they see and complex audio events they hear.
Inspired by this simple observation, an increasing number of studies expect to jointly model audio-visual information within scenes.
% to analyze and characterize from various aspects the intra-class variability and inter-class similarity of different scenes. 
% The goal is to enhance the discriminability between scenes by exploiting the complementary information from different modalities.
Recent works \cite{dcase2021_analysis}\cite{avsc2}
% \cite{dcase2021} 
show that the joint learning of acoustic and visual features can bring additional benefits to AVSC.
% To simultaneously exploit the information from both auditory and visual modalities,
To exploit the audio-visual information simultaneously,
a multi-modal system based on convolutional recurrent neural networks (CRNN) is presented in \cite{dcase2021_se}.
For better integration of 
% auditory and visual 
audio-visual
information, a multi-modal ensemble approach \cite{dcase_clip} enhanced by CLIP \cite{clip} with late fusion is proposed for AVSC.
The above AVSC systems fine-tune pretrained audio and image models on target datasets, and then fuses the intermediate representations from audio and image models, or fuses the classification decisions of audio and image models.
The decision fusion (DF) is also called late fusion, and intermediate fusion (IF) is also called early fusion \cite{dcase2021_analysis}.
% In contrast to IF, DF cannot exploit the implicit temporal and semantic correlation in audio-visual features \cite{hou21_interspeech}.
% However, no matter using IF or DF, the above AVSC systems did not fully leverage the rich and specific information of audio events and visual objects in scenes, the correlation between audio events and visual objects existing in scenes.
However, no matter using IF or DF, the above AVSC systems did not fully leverage the rich information of audio events and visual objects in scenes, as well as the correlation between them.
% audio events and visual objects.
% There are differences between audio events and visual objects in different scenes. 
% For example, in the park scene, it is common to hear birds singing and dogs barking, and to see birds on trees and dogs running by the lake. 
% In the train station scene, it is more likely to hear the train whistle, human footsteps, and speech, and then see the moving train and people passing by.
% However, the above AVSC systems do not exploit the correlation between audio events and visual objects existing in scenes.
To exploit the fine-grained information of audio events and visual objects within coarse-grained scenes, this paper proposes a contrastive learning-based alignment for audio events and visual objects to model the detailed relationship between audio-visual information.

Unlike previous studies on AVSC using IF or DF, this paper aims to exploit the fine-grained information of audio events and visual objects contained in audio-visual scenes to coordinate and fuse audio-visual modalities information for better joint modeling of the audio-visual scenes.
Therefore, this paper proposes an AVSC model equipped with contrastive event-object alignment (CEOA) and semantic-based fusion (SF).
The CEOA based on contrastive learning \cite{contrastive_leanring} aims to explore the event-object similarity within the intra-class scene and analyze the differences in the composition of audio events and visual objects between different scenes to enhance the discriminability of the model for various scenes.
The SF aims to coordinate the information of audio events and visual objects after CEOA to generate the semantic-level audio-visual information required for the final scene classification.

The contributions of this paper are: 
1) we propose contrastive learning-based event-object alignment to coordinate the relationship of fine-grained information between audio and visual modalities in scenes to assist AVSC;
2) To better fuse the audio-visual information after the alignment from CEOA, we propose SF based on cross-attention to derive the visual objects caused by audio events and the audio events caused by visual objects, and then fuse them.
3) Quantitative evaluation shows the proposed model achieves competitive performance when compared with other large-scale integrated systems, even without using additional datasets and data augmentation tricks. Visual analysis of the intermediate representations of the proposed model provides further justification for the model. 
This paper is organized as follows, Section~\ref{s2} proposes the AVSC model.
Section~\ref{s3} describes the dataset, baseline, experimental setup, and analyzes results. 
Section~\ref{s4} gives conclusions.
% an excellent trade-off between prediction performance (86.5%) and system complexity (15M parameters) 

\vspace{-0.1cm}
\section{Multi-branch AVSC model with CEOA and SF}\label{s2}

The proposed AVSC model 
% equipped with contrastive event-object alignment (CEOA) and semantic-based fusion (SF) 
in  Fig. \ref{allframeworks} consists of an audio branch, a visual branch, contrastive event-object alignment (CEOA), and semantic-based fusion (SF).
The audio branch and the visual branch generate global-level embeddings of audio events and visual objects in audio-visual clips, respectively. 
CEOA aligns embeddings
% the fine-grained information 
of audio events and visual objects in scenes to enhance the model's ability to capture the 
% detailed 
relationship between audio-visual information.
Then, semantic-level SF leverages the aligned audio-visual representations by applying cross-sensory attention and fusing the bi-modal event-object information to classify the scene.
% make the final prediction.

% \label{ssec:allframeworks}
% \begin{figure}[t]
% 	%\vspace{-0.4cm} 
% % 	\setlength{\abovecaptionskip}{0.1cm}  
%     % \setlength{\belowcaptionskip}{-6cm} 
%     % \setlength\belowcaptionskip{50\baselineskip}
%     \setlength\abovecaptionskip{-0.1\baselineskip}
% 	\centerline{\includegraphics[width = 0.75 \textwidth, angle=90]{modelv2_ok.pdf}}
% 	\caption{The proposed multi-branch AVSC model with CEOA and SF.}
% 	\label{allframeworks}
% \end{figure}

\vspace{-0.2cm}
\subsection{The audio branch}
\begin{figure}[!t]
	%\vspace{-0.4cm} 
% 	\setlength{\abovecaptionskip}{-0.1cm}  
    % \setlength{\belowcaptionskip}{-6cm} 
    % \setlength\belowcaptionskip{50\baselineskip}
    \setlength\abovecaptionskip{-0.2\baselineskip}
	\centerline{\includegraphics[width = 0.44 \textwidth]{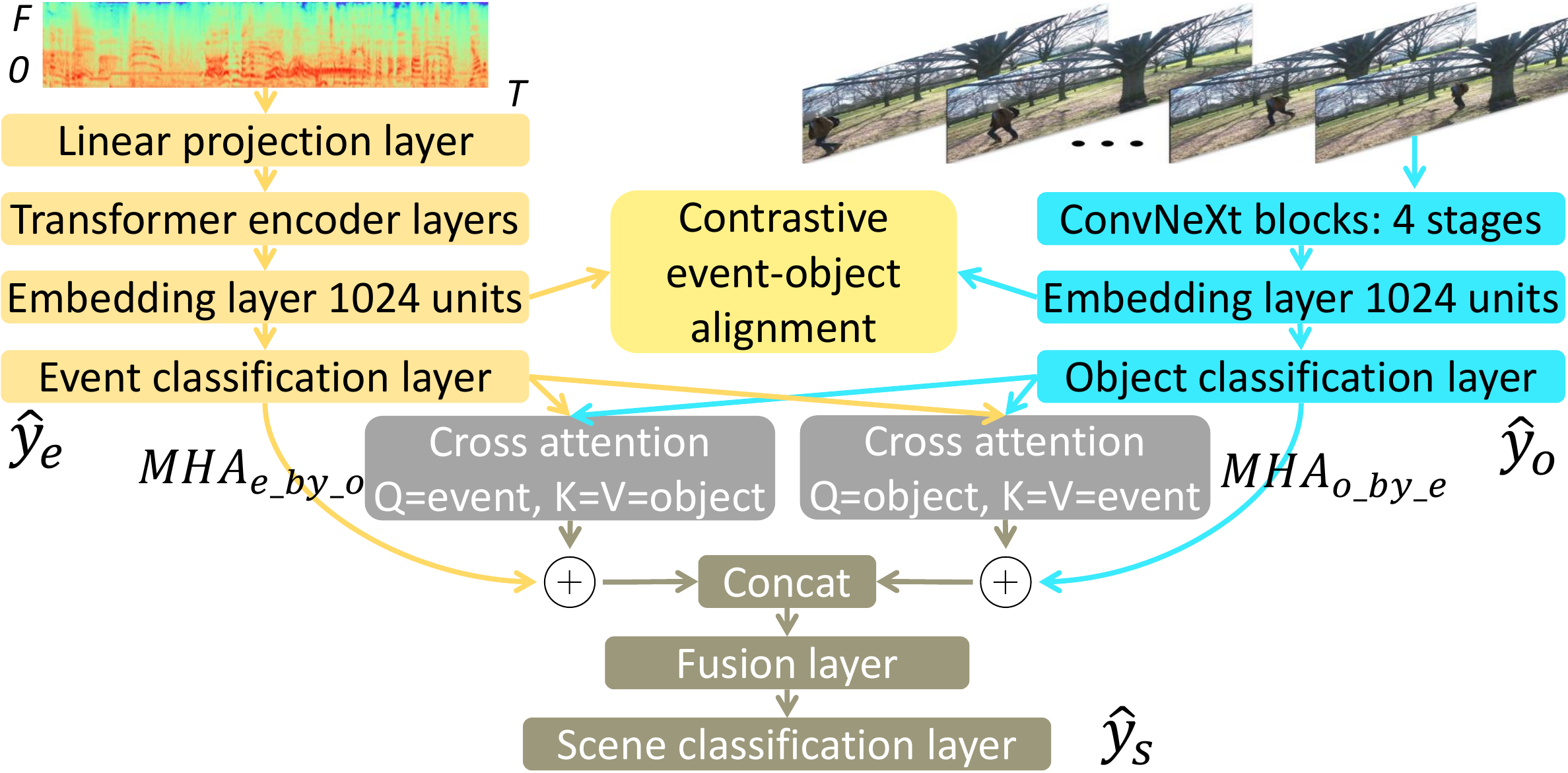}}
	\caption{The proposed multi-branch AVSC model with CEOA and SF.}
	\label{allframeworks}
\end{figure}

% The audio branch aims to learn embeddings of various audio events from audio clips.
The structure of the audio branch evolved from the Transformer \cite{transformer}, 
% with its detailed design mainly referring to
more specifically an 
Audio Spectrogram Transformer (AST) \cite{ast}.
% , which has achieved a competitive performance on audio classification tasks  \cite{audioset}. 
% on Audioset \cite{audioset}. 
% Moreover, such 
Convolution-free 
% purely attention-based 
AST can be applied to audio spectrograms and is able to capture long-range global context information \cite{ast}.
The input of the audio branch is the log-mel spectrograms \cite{mel} of a whole audio clip, and the output are probabilities of audio events that may be contained in this audio clip.
The spectrograms containing acoustic features are split into a sequence of patches. 
% Then, 
Each patch is flattened and projected onto a lower dimensional embedding space via a linear projection layer. 
% Since the patch sequence does not keep the temporal order information of the input \cite{ast}, each patch embedding is added a trainable positional embedding to allow the model to preserve temporal order in spectrograms.
Referring to AST \cite{ast}, the total number of Transformer encoder layers in Fig. \ref{allframeworks} is 12, and each layer has 12 heads for multi-head attention (MHA) \cite{transformer}.
The dimension of embedding in MHA is 768 and the dimension of each head is 64 ($768/12 = 64$), which are the same as those in \cite{touvron2021training}.
The encoder layers are followed by a linear embedding layer with ReLU activation that maps the high-level representations of audio events to labels for classification.
% For details and source code, please visit our homepage.
% As the audio branch performs multi-label classification on audio clips,
As the audio branch performs multi-label classification,
binary cross-entropy (BCE) loss is used \cite{RGASC}.
% as the loss function \cite{RGASC}. 
% More specifically, 
Denote the output of audio branch as $\hat{y}_{e} \in\mathbb{R}^{C_{e}}$, and the corresponding label as $y_e \in\mathbb{R}^{C_{e}}$, the loss can be defined as:
\begin{equation}\label{event_loss}
\setlength{\abovedisplayskip}{2pt}
\setlength{\belowdisplayskip}{2pt}
\mathcal{L}_{e}= - \sum\nolimits_{i=1}^{C_{e}} 
{y}_{e_i}\log(\hat{y}_{e_i}) + (1-{y}_{e_i})\log(1-\hat{y}_{e_i})
\end{equation}
where $C_{e}$ is the total number of event classes in the dataset, 
% $\hat{y}_{e_i}\in [0,1]$ is the probability of the occurrence of the $i$-th event in the audio clip, and ${y}_{e_i}\in \{0,1\}$ is the corresponding label.
$\hat{y}_{e_i}\in [0,1]$ is the occurrence probability of the $i$-th event in the audio clip, and ${y}_{e_i}\in \{0,1\}$ is the corresponding label.

\label{ssec:allframeworks}

\vspace{-0.2cm}
\subsection{The visual branch}
% Similar to the audio branch, the goal of the visual branch is to learn representations of visual objects from images.
Motivated by the superior performance of convolutional neural networks (CNN) in image processing, the structure of the visual branch is referred to as ConvNeXt \cite{convnext}, which is a recent version of CNN that utilizes the key components that made Transformers work well.
ConvNeXt uses depth-separable convolution, inverse bottleneck layer, Gaussian error linear unit GELU \cite{convnext}, and a larger convolution kernel (7x7) to increase the receptive field 
% of convolution 
to extract richer features.
% ConvNeXt is simple, efficient, and outperforms Transformer-based models on several image processing tasks \cite{convnext}.
% To capture the contextual and dynamic features between the image frames of a audio-visual scene stream, a sequence of frames from the stream is used as input to the visual branch.
To capture the contextual and dynamic features between images of an audio-visual scene stream, the image sequence is used as input to the visual branch.
The output of the visual branch are the probabilities of the target objects that might be contained in the image sequence.
Referring to ConvNeXt-Base \cite{convnext}, the visual branch contains 4 stages, the number of ConvNeXt blocks of 4 stages are (3, 3, 27, 3), respectively, the number of convolution channels of 4 stages is (128, 256, 512, 1024), where the number of channels doubles at each new stage.
After ConvNeXt blocks, a linear embedding layer with ReLU activation maps the high-level representations of visual objects to labels.
As the visual branch also performs multi-label classification,
% on image sequences, 
BCE loss is used. Denote the output of the visual branch as $\hat{y}_{o}  \in\mathbb{R}^{C_{o}}$, and the corresponding visual object label as $y_o \in\mathbb{R}^{C_{o}}$, the loss of the visual branch can be defined as:
\begin{equation}\label{object_loss}
\setlength{\abovedisplayskip}{1pt}
\setlength{\belowdisplayskip}{1pt}
\mathcal{L}_{o}= - \sum\nolimits_{i=1}^{C_{o}} 
{y}_{o_i}\log(\hat{y}_{o_i}) + (1-{y}_{o_i})\log(1-\hat{y}_{o_i})
\end{equation}
where $C_{o}$ is the total number of object classes in the dataset, 
$\hat{y}_{o_i}\in [0,1]$ 
% i \in [1,C_{o}]$  
is the occurrence probability of the $i$-th object in the image sequence, 
${y}_{o_i}\in \{0,1\}$ is the corresponding label.

% \vspace{-0.2cm}
\subsection{Contrastive event-object alignment (CEOA)}

\vspace{-0.02cm}
% The goal of CEOA is 
CEOA aims
to model the relationship between audio and visual modal information, and to align the fine-grained audio-visual event-object information.
% based on the distances between event-object embedding pairs.
% in different scenes.
Due to the complexity of real-life scenes, explicit event-object correlation in diverse scenes is often unknown from the input data. 
% So 
This paper expects to learn the relative distances of event-object pairs in different scenes indirectly through contrastive learning \cite{contrastive2}.
% \cite{contrastive3}, so as to 
% to characterize the event-object correlation in different scenes.
% Furthermore, the CEOA in this paper is based on pairwise contrastive loss, where the goal of contrastive loss is that like attracts like and unlike repels like.
% CL belongs to representation learning, which is essentially a method of learning a representation of samples. The target of contrastive learning is to make the representations or embeddings corresponding to similar samples closer together, and the representations or embeddings of dissimilar samples farther apart from each other.
CEOA tries to embed and coordinate representations of audio events and corresponding visual objects into the same area of the latent space so that they can be aligned in the semantic space for cross-modal fusion in the fusion part of the model.

% same area of the latent space

% More specifically, 
CEOA adopts the pairwise contrastive loss (PCL). The goal of PCL is to make the representations corresponding to positively correlated samples closer together, and the representations of samples that are less or negatively correlated farther apart.
% from each other.
% In modeling the audio-visual scenes, 
% Specifically, 
The PCL is guided by the gap between 
% anchor-positive pairs and anchor-negative 
positive-positive (PP) and positive-negative (PN) 
pairs \cite{anchor_positive}. 
% During the optimization of PCL, the consistency of the embeddings of audio events and visual objects will be gradually improved. In other words, 
During the optimization of PCL, the model will automatically focus on expanding the distance between the PP pairs and PN pairs, so as to cluster embeddings in the positive pairs and align the corresponding embeddings of audio events and visual objects to achieve fine-grained alignment of audio-visual information.
% 下面都写错了，应该是weight作为knowledge，而不是embedding是
Given $P_e \in \mathbb{R}^{K \times 1024}$ and $N_e \in \mathbb{R}^{K \times 1024}$ are the weights of $K$ audio events with the highest and lowest probabilities from the last classification layer in audio branch.
% , respectively. 
$P_o \in \mathbb{R}^{K \times 1024} $ and $N_o \in \mathbb{R}^{K \times 1024}$ are the weights of $K$ visual objects with the highest and lowest probabilities of the same input sample from the classification layer in visual branch.
These weight matrices in final classification layers can be viewed as the core
knowledge about targets learned by the model.
% For PP and PN pairs 
For the audio branch, 
% $P_e$ is used as positive of audio events, $P_o$ and $N_o$ are the corresponding positive and negative of visual objects, respectively, 
$P_e$ is used as positive samples, $P_o$ and $N_o$ are the corresponding positive and negative samples, respectively. 
% to jointly construct the PP pair $P_eP_o^T$ and PN pair $P_eN_o^T$.
% Next, 
% the event-to-object contrastive loss can be defined as: 
% Then the event-to-object contrastive loss is based on PP pair $P_eP_o^T$ and PN pair $P_eN_o^T$. 
to jointly construct the event-to-object contrastive loss by PP pair $P_eP_o^T$ and PN pair $P_eN_o^T$. 
\begin{equation}
\setlength{\abovedisplayskip}{2pt}
\setlength{\belowdisplayskip}{2pt}
\begin{aligned}
\mathcal{L}_{e2o}
&= -\ln (\text{mean}(\frac{e^{P_eP_o^T}}{e^{P_eP_o^T}+e^{P_eN_o^T}}))\\
% &= -\ln (\text{mean}(\frac{1}{1+e^{-(P_eP_o^T-P_eN_o^T)}}))\\
% &= -\ln (\text{mean}(1/({1+e^{-(P_eP_o^T-P_eN_o^T)})}))\\
&=-\ln (\text{mean}(\sigma(P_eP_o^T-P_eN_o^T)) 
\end{aligned} 
\end{equation} 
where $\sigma$ is the logistic sigmoid: $\sigma(x) = 1/({1+e^{-x}})$.
% \begin{equation}
% \setlength{\abovedisplayskip}{1pt}
% \setlength{\belowdisplayskip}{1pt}
% % \sigma(x) = \frac{1}{1+e^{-x}}
% \sigma(x) = 1/({1+e^{-x}})
% \end{equation}
% For PP and PN pairs in visual branch, $P_o$ is used as positive of objects, $P_e$ and $N_e$ are the corresponding positive and negative of events,
% , respectively, to jointly construct the PP pair $P_oP_e^T$ and PN pair $P_oN_e^T$.
% Next, the object-to-evet contrastive loss can be defined as: 
% For PP and PN pairs in 
For the visual branch, $P_o$ is used as positive samples, $P_e$ and $N_e$ are the corresponding positive and negative samples,
to build the object-to-evet contrastive loss:
\begin{equation}
\setlength{\abovedisplayskip}{2pt}
\setlength{\belowdisplayskip}{2pt}
\begin{aligned}
\mathcal{L}_{o2e}=-\ln (\text{mean}(\sigma(P_oP_e^T-P_oN_e^T)) 
\end{aligned} 
\end{equation}

% The above-mentioned 
The method of composing negative samples based on $K$ components with the lowest probabilities is called the lowest $K$ mode (LKM) in this paper.
In addition to LKM, random $K$ mode (RKM) can also be used.
% in the selection of negative components. 
That is, randomly select $K$ classes of events or objects from the weights that do not contain positive classes of events or objects.
% Compared with LKM, 
RKM will increase the difficulty of model learning and bring more challenges.
% to the model learning.
% Taking audio events as an
For example, the $K$ audio events with the lowest probabilities in LKM are the $K$ audio events that the audio branch is most confident with the least likely to occur in the input clip, while the confidence for the randomly selected $K$ audio events in RKM is ambiguous. 
% That means
So, $N_e$ in RKM will have larger information uncertainty (larger entropy) than $N_e$ in LKM.
Greater entropy will bring more burdens and possibilities for model learning, 
% and potentially better performance and results.
% this paper will explore the impact of $K$ and 2 modes 
% % (LKM and RKM) 
% on model performance in the experiment. 
the impact of $K$ and 2 modes on model performance will be explored in the experiments.

% \vspace{-0.1cm}
\subsection{Semantic-based fusion (SF)} 
% \vspace{0cm}
% The goal of SF is to perform cross-modal fusion of audio embeddings containing events information and visual embeddings containing objects information to generate semantic-level audio-visual representations after fine-grained alignment based on contrastive loss. 
% The goal of 

\vspace{-0.05cm}
SF aims to perform cross-modal fusion of audio and visual embeddings to generate semantic-level audio-visual representations after fine-grained alignment by CEOA. 
To 
% comprehensively 
consider the possible interactions and correlations of audio events and visual objects in diverse scenes, 
% inspired by attention in Transformer \cite{transformer}, 
% this paper proposes the semantic-level fusion module SF based on the multi-head attention (MHA). 
this paper proposes SF based on the multi-head attention (MHA) \cite{transformer}, which is calculated on a set of queries ($\mathbf{Q}$), keys ($\mathbf{K}$), and values ($\mathbf{V}$).
\begin{equation}
\setlength{\abovedisplayskip}{3pt}
\setlength{\belowdisplayskip}{3pt}
\begin{split}
MHA(\mathbf{Q,K,V})=Concat(head_{1}, ..., head_{h} )\mathbf{w}^{O} 
\\
where \quad head_{i}= A(\mathbf{Qw}_{i}^{Q}, \mathbf{Kw}_{i}^{K}, \mathbf{Vw}_{i}^{V}), \\
A(\mathbf{Qw}_{i}^{Q}, \mathbf{Kw}_{i}^{K}, \mathbf{Vw}_{i}^{V})=\Phi({\mathbf{Qw}_{i}^{Q}}{\mathbf{Kw}_{i}^{K}}^T/\sqrt{d})\mathbf{Vw}_{i}^{V}
\end{split}
\end{equation}
Referring to settings of Transformer~\cite{transformer},
$\Phi$ is softmax function,
the total number of attention heads $h$ is 8,  learnable weights $\{\mathbf{w}_{i}^{Q}, \mathbf{w}_{i}^{K}, \mathbf{w}_{i}^{V}\}\in \mathbb{R}^{1 \times d}$ and $d=64$, $\mathbf{w}^{O} \in \mathbb{R}^{(h*d) \times 1}$.
% Following settings of Transformer~\cite{transformer},
When $\mathbf{K}$ and $\mathbf{V}$ are core event embeddings from the final event classification layer ($\{\mathbf{K}, \mathbf{V}\}\in \mathbb{R}^{C_e \times 1}$), and $\mathbf{Q}$ is core object embeddings from the final object classification layer ($\mathbf{Q} \in \mathbb{R}^{C_o \times 1}$), 
the output of MHA can be viewed as visual objects embeddings caused by audio events, which is denoted as $MHA_{o\_by\_e} \in \mathbb{R}^{C_o \times 1}$.
% Among them, 
In this, 
$\mathbf{Qw}_{i}^{Q}{\mathbf{Kw}_{i}^{K}}^T \in \mathbb{R}^{C_o \times C_e}$ can be regarded as the transformation matrix from audio space to visual space. 
% after the transformation matrix, 
% $\mathbf{Vw}_{i}^{V} \in \mathbb{R}^{C_e \times d}$ represents audio information is transformed into the visual semantic space to try to couple the corresponding visual objects.
In contrast, when $\mathbf{K}$ and $\mathbf{V}$ are core object embeddings ($\{\mathbf{K}, \mathbf{V}\}\in \mathbb{R}^{C_o \times 1}$), and $\mathbf{Q}$ is core event embeddings ($\mathbf{Q} \in \mathbb{R}^{C_e \times 1}$), 
the output of MHA can be viewed as audio events embeddings caused by visual objects, which is denoted as $MHA_{e\_by\_o} \in \mathbb{R}^{C_e \times 1}$.
Next, as shown in Fig. \ref{allframeworks}, $MHA_{o\_by\_e}$ is added with object embeddings $\hat{y}_{o}$ 
% output by the visual branch 
to produce audio-visual enriched objects embeddings, $MHA_{e\_by\_o}$ is added with event embeddings $\hat{y}_{e}$ 
% output by the audio branch 
to produce audio-visual enriched events embeddings.
Then, the event and object embeddings are concatenated together to form audio-visual semantic embeddings, and the fusion layer with ReLU activation 
% coordinates and 
maps the audio-visual embeddings into scene classes.
% Scene classification is usually considered as single-label multi-class classification problem, 
As scene classification performs single-label multi-class classification,
cross-entropy loss \cite{RGASC} is used between the output 
% of the proposed AVSC model 
$\hat{y}_{s}\in\mathbb{R}^{C_s}$ and the scene label ${y}_{s}\in\mathbb{R}^{C_s}$,
\begin{equation}
\setlength{\abovedisplayskip}{2pt}
\setlength{\belowdisplayskip}{2pt}
\mathcal{L}_{s} = - \sum\nolimits_{i=1}^{C_s} 
{y}_{s_i}\log(\hat{y}_{s_i})
\end{equation}
where $C_{s}$ is the total number of scene classes in the dataset, 
$\hat{y}_{s_i}\in [0,1]$ 
% i \in [1,C_{o}]$. 
% $\hat{y}_{s_i}$ 
is the occurrence probability of the $i$-th 
% audio-visual 
scene in the input clip, and ${y}_{s_i}\in \{0,1\}$ is the corresponding label.

In the training phase, to help the model comprehensively consider the fine-grained information of audio events ($\mathcal{L}_{e}$), visual objects ($\mathcal{L}_{o}$), contrastive event-to-object ($\mathcal{L}_{e2o}$), contrastive object-to-event ($\mathcal{L}_{o2e}$), and coarse-grained global information of audio-visual scene ($\mathcal{L}_{s}$) within the input clips, the final loss of the proposed model
% equipped with CEOA and SF 
can be defined as:
%\begin{equation*}%加*表示不对公式编号
\begin{equation}
	\setlength{\abovedisplayskip}{2pt}
	\setlength{\belowdisplayskip}{2pt}
	\begin{split}
	\mathcal{L}=&\lambda_1\mathcal{L}_{e}+\lambda_2\mathcal{L}_{o}+\lambda_3\mathcal{L}_{e2o}+\lambda_4\mathcal{L}_{o2e}+\lambda_5\mathcal{L}_{s} 
	\end{split}
\end{equation}
% where $\lambda_i$ is the scale factor of each loss function and defaults to 1. 
% $\lambda_i$ determines the importance of each loss function in training.  
% In the experimental section, various configurations of $\lambda_i$ are explored. 
where $\lambda_i$ is the scale factor of each loss and defaults to 1. 
% $\lambda_i$ determines the importance of each loss in training.  
% In the experimental section, various configurations of $\lambda_i$ are explored. 

% \vspace{-0.15cm}
\section{Experiments and results}\label{s3}

% \vspace{-0.1cm}
\subsection{Dataset, Baseline, Experiments Setup, and Metrics}

\vspace{-0.1cm}
% \textbf{Dataset.}
% The audio-visual scene dataset used in this paper is 
TAU Audio-Visual Urban Scenes 2021 development dataset~\cite{dcase2021} used in this paper consists of 12291 10-seconds clips totaling 34.14 hours and contains 10 different classes of audio-visual scenes.
% These audio-visual recordings were recorded in 12 different European locations. 
This real-life dataset
% \protect\footnote{Dataset available:  https://zenodo.org/record/4477542\#.YqsEcOhBxnI} 
does not contain labels for audio events nor visual objects. 
% Therefore, 
% To obtain labels of events and objects for model training, 
% Transformer-based 
% AST\protect\footnote{Pre-trained AST model available: \url{https://www.dropbox.com/s/ca0b1v2nlxzyeb4/audioset_10_10_0.4593.pth?dl=1}} \cite{ast} and 
AST\protect\footnote{AST with 0.459 mAP: \url{https://github.com/YuanGongND/ast}} \cite{ast} and 
% CNN-based 
% ConvNeXt\protect\footnote{Pre-trained ConvNeXt model available: \url{https://download.pytorch.org/models/convnext_base-6075fbad.pth}} 
ConvNeXt\protect\footnote{\url{https://download.pytorch.org/models/convnext_base-6075fbad.pth}} 
\cite{convnext} are used to tag each clip with pseudo labels indicating the probability of audio events and visual objects for model training.
% , respectively. 
% Since AST is trained on Audioset \cite{audioset} with 527 classes of audio events and ConvNeXt is trained on ImageNet-1K \cite{imagenet} with 1000 classes of visual objects, the number of classes of events and objects in pseudo labels is 527 and 1000.
Since AST and ConvNeXt are trained on Audioset \cite{audioset} (527 classes) and ImageNet-1K \cite{imagenet} (1000 classes), respectively, the number of classes of audio events and visual objects in pseudo labels is 527 and 1000.
% However, in a 10-second clip from real-life scenes, it is almost impossible to have 527 class of audio events and 1000 class of visual objects at the same time. 
% To binarize the probabilities of events and objects output by pre-trained models into hard labels consisted of 1 and 0, the thresholds for the occurrence of audio events and visual objects need to be set,
% in the experiment, 
% respectively.
% If the 
% % occurrence 
% threshold 
% % for audio events 
% is small, the label vector of some 
% % audio 
% clips will be full of noise. 
% However, if the threshold is large, the label vector of some 
% % audio 
% clips will be all 0.
% Therefore, to reduce the noise in hard labels and make all clips contain valid labels, this paper selects 0.0365 and 0.9216 as the occurrence threshold for audio events and visual objects, respectively, which results in the number of classes of events and objects used in this paper being 306 and 704.
% , that is, $C_{e}$ in Equation (Eq.) \ref{event_loss} is 306.
% Similarly, 0.9216 is used as the occurrence threshold for visual objects, which results in the number of visual object classes in this paper being 704, that is, $C_{o}$ in Eq. \ref{object_loss} is 704.
% This paper uses 0.0365 and 0.9216 as the occurrence probability threshold for audio events and visual objects, respectively, to binarize the probabilities of events and objects output by pre-trained models into hard labels consisted of 1 and 0. 
We set the thresholds (0.0365 and 0.9216) for the occurrence of audio events and visual objects to binarize the probabilities output by pre-trained models into hard labels consisting of 1 and 0. 
% This paper uses 0.0365 and 0.9216 as the occurrence probability threshold for audio events and visual objects, respectively, to binarize the probabilities of events and objects output by pre-trained models into hard labels consisted of 1 and 0. 

To compare the performance of the model with other models on the same benchmark, this paper uses the baseline in DCASE 2021 Task 1 Subtask 
% B\protect\footnote{DCASE 2021 Task 1 Subtask B:~\url{https://dcase.community/challenge2021/task-acoustic-scene-classification}} 
B
% \protect\footnote{\url{https://dcase.community/challenge2021/task-acoustic-scene-classification}} 
(T1B)~\cite{dcase2021_analysis} as baseline, and further compares the proposed model with other methods from different perspectives on the same AVSC task.
% with contrastive learning and semantic-based attention fusion.
To facilitate the comparison with other methods,
the training/testing split of the dataset follows the default split of T1B.
Similar to all other comparison methods using weights from pre-trained models, 
the proposed model also uses part of the weights from AST \cite{ast} and ConvNeXt \cite{convnext} during training.

For training, log mel-bank energy with 128 banks \cite{mel} is used as acoustic features,
% in the audio branch, 
which is extracted by STFT with Hamming window length of 25~\textit{ms} and a hop size of 10~\textit{ms} between the window. 
% then the spectrogram is split into a sequence of patches following settings of \cite{ast}.
% To 
% % comprehensively consider 
% exploit the contextual and dynamic information of images, the 
An image sequence consisting of one image per second is input to the visual branch. 
% That is, 
For a 10-second scene clip, the input to the visual branch is an image sequence consisting of 10 images.
A batch size of 16 and AdamW optimizer \cite{adamw} with learning rate of 5e-6 are used to minimize the losses in the proposed model. 
To prevent over-fitting, dropout \cite{dropout} and normalization are used.
Systems are trained on a single card Tesla V100-SXM2-32GB for 100 epochs.
The Logloss and average 
% class-wise 
accuracy (Acc.) \cite{dcase_kong} are used as metrics. 
Larger Acc. and lower Logloss indicate better performance.  
% For more details and the source code,
% please visit the project 
More details,
please see the homepage\protect\footnote{Homepage: \url{https://github.com/Yuanbo2020/Contrastive-AVSC}}. %
% More details and code,
% please see the homepage\protect\footnote{Homepage: \url{https://github.com/Yuanbo2020/Contrastive-AVSC}}. % (\textcolor{blue}{\underline{\url{https://github.com/Yuanbo2020/Contrastive-AVSC}}}).
% More details and code,
% please see the homepage (\url{https://github.com/Yuanbo2020/Contrastive-AVSC}).

% \vspace{-0.1cm}
\subsection{Results and Analysis}
% \vspace{-0.12cm}
\textbf{Difference between LKM and RKM.}
% In this paper, the negative samples in CEOA have two modes when selecting components, 
There are 2 modes in CEOA when selecting negative samples,
LKM based on the lowest occurrence probabilities and RKM based on random screening.
Compared with LKM, RKM increases the learning difficulty of the model for fine-grained 
% specific 
event-object pairs due to the greater uncertainty of negative components.
% Therefore, 
Table~\ref{tab:topk} shows the effects of the two modes on model performance.
% to choose the final negative samples selection mode used in this paper.

% \tiny
% \scriptsize
% \footnotesize
% \small
% \normalsize
% \large
% \Large
% \LARGE
% \huge
% \Huge
% \vspace{-0.45cm}

% \vspace{-0.3cm}
% As shown in Table~\ref{tab:topk}, 
As the value of $K$ increases, the classification accuracy of the model in Table~\ref{tab:topk} in both modes improves. That is, the gradual increase in the influence of contrastive learning in model training is beneficial to the model's recognition of fine-grained event-object pairs in audio-visual scenes, which in turn helps to improve the model's ability to identify differences between different scenes.
% However, 
When $K$ increases to a certain value, increasing the value of $K$ will not bring more benefits to the scene analysis ability of the model. 
% That means after the components used in contrastive learning reach a certain size, increasing the size of the components will not improve the scene classification effect of the model.
The models in LKM and RKM 
% in Table~\ref{tab:topk} 
achieve the best performance when $K$ is 15 and 10, respectively,  where the performance of models in RKM are slightly inferior to that in LKM. 
The reason may be that the randomness of negative components in RKM brings more challenges and difficulties to the learning of the model, making it difficult for the model to achieve a balance between extracting and coordinating audio-visual event-object representations and efficiently capturing information for scene classification.  
However, it is worth noting that in LKM, varying $K$ values obviously affects the model performance, while different $K$ values in RKM do not have a much different impact on the model. 
This may be due to the fact that the negative components in event-object pairs of RKM are randomly selected, which makes the model less sensitive to the size of the contrastive pairs in aligning event-object representations.
Based on the results in Table 2,
% with different $K$ values for the two models,
LKM and $K=15$ will be used in subsequent experiments.

\vspace{-0.3cm}
\begin{table}[h]\footnotesize
	% 表格标题的距离 above设置标题上面的距离，below设置标题下面的距离
	\setlength{\abovecaptionskip}{-0.1cm}   %表格和标题之间的距离
	\renewcommand\tabcolsep{1pt} %来调整表格的列间距离
	\centering
	\caption{Acc. (\%) of the proposed AVSC model in different modes.}
	\begin{tabular}
	{
	p{1cm}<{\centering} |
	p{0.8cm}<{\centering} 
	p{0.8cm}<{\centering}
	p{0.8cm}<{\centering}
	p{0.8cm}<{\centering}
	p{0.8cm}<{\centering}
	p{0.8cm}<{\centering}
	p{0.7cm}<{\centering}   
	} 
		%\toprule[1pt] %最上面表格的粗细
		%\specialrule{0em}{0pt}{0pt}
		%specialrule 命令第一个大括号控制表格线的粗细，若为0，则表格线透明，第二个大括号是表格线与上方内容的距离，第三个大括号是表格线与下方内容的距离，通过改变后两个大括号中的值来控制行高！
		   
		\toprule[1pt]
		\specialrule{0em}{0pt}{0pt}
		%\hline
		
		\textit{K value}
		& \textsl{1} 
		& \textsl{5} 
		& \textsl{10} 
		& \textsl{15} 
		& \textsl{20} 
		& \textsl{25} 
		& \textsl{30}   \\
		
% 		\cline{1-8}
        \hline
		
		\textit{LKM}
		& \textit{89.32}  & 
		\textit{90.12}  & 
		\textit{90.83} & 
		\textit{\textbf{91.58}} & 
		\textit{91.38} & 
		\textit{91.00} &  \textit{90.91} \\
		
		\hline
		 
		\textit{RKM} 
		& \textit{91.00} & 
		\textit{91.11} & 
		\textit{\textbf{91.30}} & 
		\textit{91.27} & 
		\textit{91.22} & 
		\textit{91.08} &  \textit{91.05} \\
		
		\specialrule{0em}{0pt}{0pt}
		\bottomrule[1pt]
		
	\end{tabular}
	\label{tab:topk}
\end{table}

% \vspace{-0.2cm}
\textbf{Gain of adding CEOA and SF to the model.}
The proposed
% multi-branch 
AVSC model 
% equipped with CEOA and SF 
% proposed in this paper 
aims to use the contrastive learning-based CEOA to align the fine-grained event-object information within audio-visual scenes to coordinate the relationship between the audio-visual information, and then utilize the attention-based SF to collaboratively fuse audio-visual representations across modalities.
% Table~\ref{tab:module} reveals the gain of the proposed modules for the model learning capability in this paper.
% Please note that in the multi-branch backbone without SF, simple concatenation is used to fuse the audio-visual modalities information.
Table~\ref{tab:module} summarizes the gain of the proposed modules for the model learning capability.
% in this paper,
% in \# 1, simple concatenation is used to fuse the audio-visual information.
Compared to the basic backbone, both the contrast learning-based CEOA and the semantic-level attention-based fusion proposed in this paper improves the classification ability of the model for audio-visual scenes. 
Among them, CEOA 
% from the perspective of aligning fine-grained event-object information to coordinate the relationship between audio and visual information 
has a slightly larger improvement on the model performance. 
This also illustrates the benefit of mining the fine-grained event-object information contained in diverse scenes for recognition. Finally, the backbone equipped with CEOA and SF achieves better results.
% after performing the attention-based cross-modal fusion of the audio-visual information aligned by contrastive learning.

\vspace{-0.4cm}
\begin{table}[h]\scriptsize % footnotesize   % \normalsize
	% 表格标题的距离 above设置标题上面的距离，below设置标题下面的距离
	\setlength{\abovecaptionskip}{-0.1cm}   %表格和标题之间的距离
	\renewcommand\tabcolsep{4pt} 
	\centering
	\caption{The ablation study of the proposed modules.}
	\begin{tabular}
	{p{0.6cm}<{\centering}|
	p{1.2cm}<{\centering}
	p{1.2cm}<{\centering}
	p{1.2cm}<{\centering}|
	p{1.2cm}<{\centering}
	p{1.2cm}<{\centering}}
	
    \hline 
	   {\#} & \textit{Backbone} 
	   & \textit{SF}
	   & \textit{CEOA}
	   & \textsl{Acc. (\%)} 
	   & \textsl{Logloss} \\
		\hline  % 中间的横线粗细
		\specialrule{0em}{0.05em}{0.pt}
		1 & \CheckmarkBold & \XSolidBrush  & \XSolidBrush  &  88.42  & 0.439 \\ 
		2 &  \CheckmarkBold & \CheckmarkBold  & \XSolidBrush  &  90.34 & 0.390  \\
		3 &  \CheckmarkBold & \XSolidBrush  & \CheckmarkBold  &  90.75 & 0.357  \\
		4 &  \CheckmarkBold & \CheckmarkBold  & \CheckmarkBold  & \textbf{91.58} & \textbf{0.259}   \\
	\hline
	\end{tabular}
	\label{tab:module}
\end{table}

\vspace{-0.2cm}
\textbf{Weighting scale factors in the loss.}
Audio-visual scene naturally contains both audio and visual information, and the proposed CEOA also contains two kinds of contrastive information from the perspectives of audio events and visual objects. 
% The question of whether different types and different modalities of information should be given equal importance during the training phase needs to be further explored.
In training, different coefficients of loss components represent the importance of their corresponding targets in the overall model performance. 
% In addition, 
Different combinations of coefficients often imply different concerns of the model in the learning process.
% Therefore, the next step is to investigate the ratio of coefficients between different components of loss in the 
% % multi-branch 
% proposed multi-loss model to optimize the utilization of diverse information in the training.
% Table~\ref{tab:lambda} shows the performance of the model with several combinations of loss weights. 
Table~\ref{tab:lambda} summarizes the model performance with several combinations of loss weights. It shows the effect of changing the ratio between different components of the loss on the utilization of diverse information in the training.

\vspace{-0.4cm}
\begin{table}[h]\scriptsize %scriptsize % small % footnotesize
	% 表格标题的距离 above设置标题上面的距离，below设置标题下面的距离
	\setlength{\abovecaptionskip}{-0.1cm}   %表格和标题之间的距离
	\renewcommand\tabcolsep{1.5pt} 
	\centering 
\caption{The effect of different $\lambda_i$ values on the AVSC task.}
	\begin{tabular}
	{
	p{0.6cm}<{\centering}|
	p{0.85cm}<{\centering}
	p{0.85cm}<{\centering}
	p{0.85cm}<{\centering}
	p{0.85cm}<{\centering}
	p{0.85cm}<{\centering}|
	p{1.4cm}<{\centering}
	p{1.4cm}<{\centering}
	}  
		\hline 
		{\#} & 
		$\lambda_1$ &  
		$\lambda_2$ & 
		$\lambda_3$ &  
		$\lambda_4$ &  
		$\lambda_5$ &
		\textsl{Acc. (\%)} & 
		\textsl{Logloss} \\ 
\hline
     1 & 1 &  1 &  1 &  1  & 1  & 91.58 & 0.259 \\ 
     2 & 0.5 &  1 &  1 &  1  & 1  &  91.82 & 0.258  \\

     3 & 0.5 &  0.5 &  1 &  1  & 1  & 92.20  &  0.237 \\

     4 & 0.25 &  0.5 &  1 &  1  & 1  & 93.06  & 0.226  \\

     5 & 0.25 &  0.5 &  0.25 &  1  & 1  & 93.71  &  0.222 \\

     6 & 0.25 &  0.5 &  0.25 &  0.5  & 1  & \textbf{94.10}  & \textbf{0.192}  \\
     7 & 0.25 &  0.5 &  0.01 &  0.01  & 1  & 93.99  &  0.193 \\

     8 & 0.05 &  0.25 &  0.05 &  0.25  & 1  & 93.44  & 0.254  \\

     9 & 0.01 &  0.5 &  0.01 &  0.01  & 1  & 93.52  & 0.209  \\ 

     10 & 0.01 & 0.1 &  0.01 &  0.1  & 1  & 93.33  & 0.246  \\

     11 & 0.01 & 0.01 &  0.05 &  0.1  & 1  & 94.02  & 0.193  \\

     12 & 0.005 &  0.1 &  0.025 &  0.1  & 1  & 93.41  & 0.219  \\

\hline

	\end{tabular}
	\label{tab:lambda}
\end{table}

\vspace{-0.3cm}
% Table~\ref{tab:lambda} explores fine-grained fine-tuning with different weights for multiple loss components. 
The fusion of cross-modal information of different types in Table~\ref{tab:lambda} aims to maintain as much as possible the recognition ability of the audio model and visual model for audio events and visual objects, respectively, while making full use of the implicit relationship between audio events and visual objects, so that they both contribute to the model's ability of classifying scenes. 
% Therefore, 
Several combinations of coefficients are explored 
% in Table~\ref{tab:lambda} 
to select the optimal ratio between the information of audio events, visual objects, event-object pairs and audio-visual scenes for a better AVSC model.
Finally, giving maximum weight to $\mathcal{L}_{s}$ and secondary weight to the information related to visual objects ($\mathcal{L}_{o}$, $\mathcal{L}_{o2e}$), while absorbing the information related to audio events ($\mathcal{L}_{e}$, $\mathcal{L}_{e2o}$), makes the best result of \# 6 in Table~\ref{tab:lambda}.
This combination of coefficients gives a stronger emphasis on the visual object information, which means that the visual object information is more important than audio event information in the proposed model for the AVSC task.

% alpha[0] * loss_event + alpha[1] * loss_object + alpha[2] * loss_scene \
                            %   + alpha[3] * contrastive_loss_auido + alpha[4] * contrastive_loss_image

% Le + λ2Lo + λ3Le2o + λ4Lo2e + λ5Ls     

\vspace{-0.2cm}
\begin{table}[!h]\scriptsize % footnotesize   % \normalsize
	% 表格标题的距离 above设置标题上面的距离，below设置标题下面的距离
	\setlength{\abovecaptionskip}{-0.1cm}   %表格和标题之间的距离
	\setlength{\belowcaptionskip}{-0.47cm}  
	\renewcommand\tabcolsep{4pt} 
	\centering
	\caption{Performance of the model on different modal information.}
	\begin{tabular}
	{p{0.8cm}<{\centering}|
	p{1.3cm}<{\centering}
	p{1.3cm}<{\centering}|
	p{1.5cm}<{\centering}
	p{1.5cm}<{\centering}}
    \hline 
	   {\#} & \textit{Audio} 
	   & \textit{Visual}
	   & \textsl{Acc. (\%)} 
	   & \textsl{Logloss} \\
		\hline  % 中间的横线粗细
		\specialrule{0em}{0.05em}{0.pt}
		1 & \CheckmarkBold & \XSolidBrush    &  73.55  &  0.871  \\ 
		2 &  \XSolidBrush & \CheckmarkBold    &  88.86  &  0.518  \\
		3 &  \CheckmarkBold & \CheckmarkBold  & \textbf{94.10} & \textbf{0.192}   \\
		
% 		max_val_acc: 0.940192

	\hline
	\end{tabular}
	\label{tab:audio-visual}
\end{table}

\textbf{Comparison of single-modal and multi-modal models.}
% After coordinating the relationship of various information among the proposed AVSC model, the final AVSC model equipped with CEOA and SF is obtained. 
Compared with the single-modal audio or visual model, the audio-visual model can utilize both modal information,
% at the same time, 
and understand differences in the description of the same target from the perspectives of different modalities.
% , so as to perform cross-modal fusion and generate audio-visual semantic information about targets.
In Table~\ref{tab:audio-visual}, the audio-visual model that fuses fine-grained information from cross-modal achieves better results,
% shows the performance of the proposed AVSC model with single- and multi-modal information,
% In Table~\ref{tab:audio-visual}, 
the audio-only model performs the worst, while the result of the visual-only model is slightly better. 
This is consistent with the trend reflected in Table~\ref{tab:lambda}, that is, visual object information is more valuable than audio event information in the proposed model for the AVSC task.
The reason for this phenomenon may be that in this dataset, the audio clips between different scenes do not sound very different, and most of the audio clips are full of noise, making it difficult for even a human to effectively distinguish target scenes by relying only on audio clips.
% modal information.
However, the differences between visual objects in different scenes are obvious, a park with trees and a lake is clearly different from an airport with monitors and escalators. 
Therefore, using visual information can effectively distinguish between different scenes in this AVSC task, which also leads to the result that the visual object information plays a more important role than audio events.
% The audio-visual model can utilize the fine-grained information of audio events and visual objects, and can also exploit the coordinated relationship between the aligned bi-modal event-object information. Therefore, the audio-visual model that fuses fine-grained information from cross-modal achieves better results.
% And the audio-visual model that fuses fine-grained information from cross-modal achieves better results.

% \vspace{-0.2cm}
\textbf{Comparison with prior methods.}
% In Table~\ref{tab:other_models}, the performance of the proposed model is compared with other systems on the same AVSC task.
% The AVSC model proposed in this paper does not use any audio or image data augmentation (Aug.) methods in the training, nor does it use any ensemble methods. 
% Therefore, for a fair comparison, we extract the results of the best single model from other systems directly from the T1B competition website.
Table~\ref{tab:other_models} shows the performance of the proposed model and other systems on the same AVSC task.
The proposed model does not use any data augmentation (Aug.) methods, nor does it use any ensemble methods. 
So, for a fair comparison, we extract the best single model result of other systems directly from the T1B website.

% % 使用额外的场景数据集Places365来微调模型，并使用了RandomResizedCrop, RandomHorizontalFlip, random adjustment
% of image brightness and contrast and other data augmentation tricks.
% In addition, we use Mixup[14] data augmentation for the images
% 等数据增强方法 

Table~\ref{tab:other_models} includes published results of the top 5 teams in T1B competition, 
all systems listed in Table~\ref{tab:other_models} use the weights of pre-trained models involved.
Since the result of a single model could not be found in the paper \cite{dcase_clip} of \# 6, we implemented their system according to their settings in \cite{dcase_clip}.
Except for Baseline and the proposed model do not use data augmentation, other systems use diverse audio or visual data augmentation methods.
% , such as Pitch shifting, SpecAugment, Mixup, AutoAugment, Random Gain, Frequency Masking, RandomAffine, ColorJitter, GaussianBlur, Random Erasing, etc.
The system of \# 7 uses an additional large scene dataset Places365 for training. To compare the model performance on the same dataset ImageNet, we select the result based on ImageNet in \# 7.
The model in \# 9 is trained with both ImageNet and Places365, and also uses a two-stage fine-tuning strategy.
% that combines at least 5 different data augmentation methods.
In contrast, the proposed model, which does not involve data augmentations, achieves similar results to that of \# 9, which uses additional datasets and multiple data augmentations.
% In other words, 
That is, even without using additional datasets and data augmentation tricks, the proposed model shows competitive performance.

\vspace{-0.2cm}
\begin{table}[h]\scriptsize
	% 表格标题的距离 above设置标题上面的距离，below设置标题下面的距离
% 	\setlength{\abovecaptionskip}{0cm}   %表格和标题之间的距离
% 	\setlength{\belowcaptionskip}{-0.5cm}   %???????????????????
	\renewcommand\tabcolsep{1.5pt} 
	\centering
	\caption{Comparison of audio-visual scene classification results of different systems on the same dataset.}
	\begin{tabular}{
	p{0.4cm}<{\centering}|
	p{2.7cm}<{\centering}|
	p{1.5cm}<{\centering}|
	p{1.5cm}<{\centering}|
	p{1.0cm}<{\centering}|
	p{1.0cm}<{\centering}
	}
	    \hline
		\multirow{2}{*}{\#} & \multirow{2}{*}{System} & 
		Audio Backbone & 
		Visual Backbone & 
		Aug. Method & 
		\multirow{2}{*}{\textsl{Acc (\%)}}\\
		\hline 
		1 & Baseline \cite{dcase2021_analysis} & OpenL3 & OpenL3 & \textit{None} & 77.0\\
		
		 2& WaveTransformer \cite{wavetrans} & OpenL3 & OpenL3 & 3 types & 79.5\\
		
		3 & CRNN \cite{dcase2021_se} & SE-Net & VGG16 & 1 type & 90.0\\
		
		4 & 2-stage classifier \cite{top_bit} & fsFCNN & TimeSformer & 8 types & 91.5\\
        %  fsFCNN (frequency sub-sampling controlled fully convolution) 
        5 & 2-stream model \cite{top_hou} & OpenL3 & ResNet50 & 3 types & 91.7\\
        
		6 & CLIP variants \cite{dcase_clip} & EfficientNet & CLIP ViT & 6 types & 93.3\\
		% 由于没有开源，所以我们根据论文中的设定自己实验了一下
		
		7 & AVSM \cite{top2} & VGGish & ResNet50 & 8 types & 93.5\\
		% 为了在同一个基准上对比，我们选取了方法2中与本文一样同样使用ImageNet训练的结果。
		
		8 & CNN\_Transformer \cite{top4} & VGGish & Transformer & 4 types & 93.9\\
	    
	   9 & 2-stage fine-tuning \cite{top3} & ResNet & EfficientNet & 5 types & \textbf{94.1}\\
	    % 使用额外的场景数据集Places365来微调模型，并使用了RandomResizedCrop, RandomHorizontalFlip, random adjustment
% of image brightness and contrast and other data augmentation tricks.
% In addition, we use Mixup[14] data augmentation for the images
% 等数据增强方法 
        10 & Proposed model & Transformer & ConvNeXt & \textit{None} & \textbf{94.1} \\  
	\hline
	\end{tabular}
	\label{tab:other_models}
\end{table}

% That is, even without using additional datasets and data augmentation tricks, the proposed model, which is accompanied by contrastive learning-based audio-visual event-object alignment and multi-headed attention-based cross-modal fusion, shows competitive performance.

\begin{figure}[!t]
	%\vspace{-0.4cm} 
 \setlength\abovecaptionskip{-0.2\baselineskip}
	\centerline{\includegraphics[width = 0.47 \textwidth]{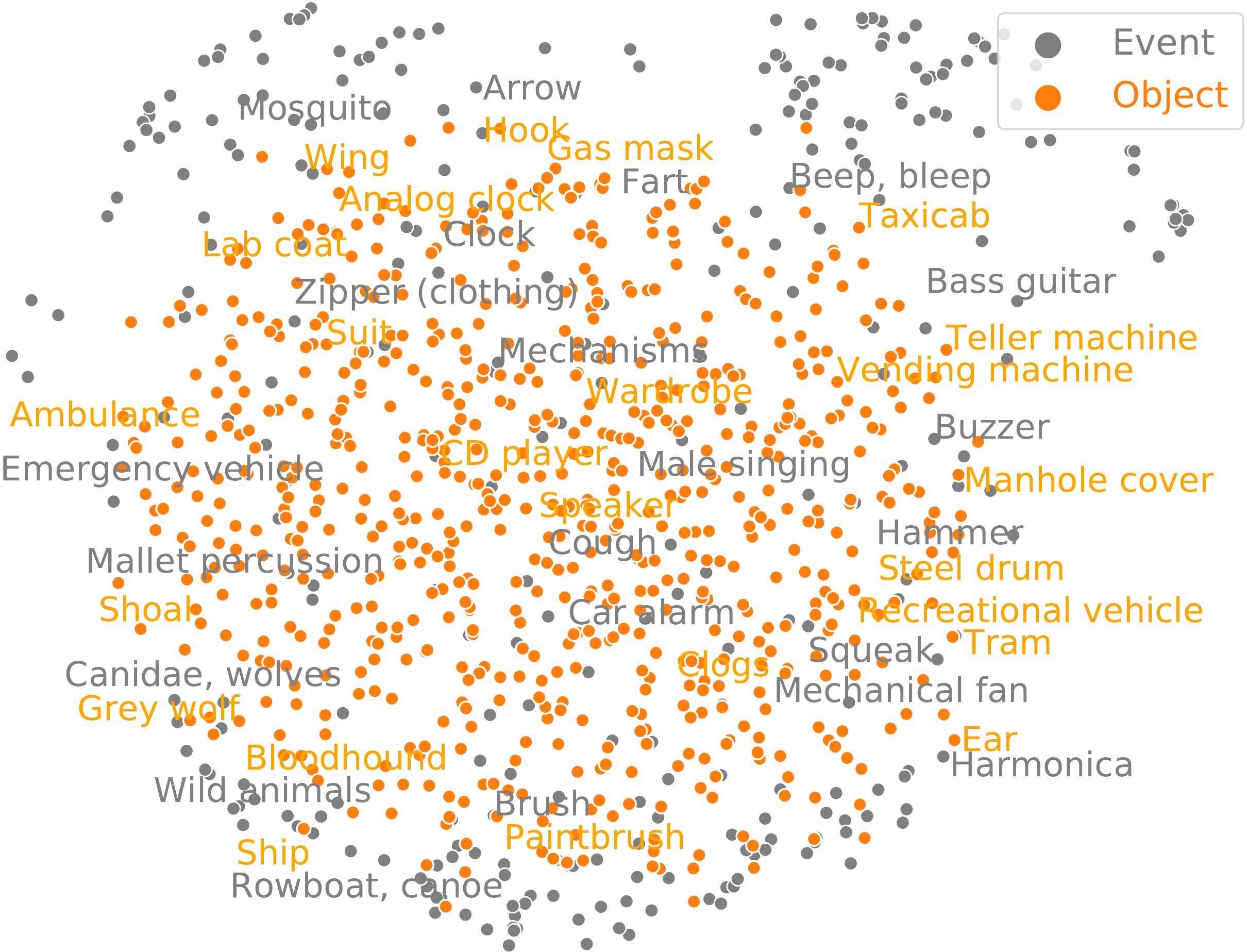}}
	\caption{Visualization of the learned weights of events and objects from the event classification layer and object classification layer, respectively.
	Please visit the homepage to interactively view the label for each point in detail.}
	\label{w_s_e}
\end{figure}

\vspace{-0.2cm}
\textbf{Visual analysis.}
To gain concise insights into the event-object cross-modal alignment, Fig.~\ref{w_s_e} intuitively visualizes the core knowledge (weights from the classification layers) about audio events and visual objects learned by the model with CEOA using UMAP \cite{umap}.
Different kinds of audio events and visual objects are interleaved and regularly distributed in Fig.~\ref{w_s_e}. 
% Emergency vehicle (event) and Ambulance (object), 
% Grey wolf (event) and Canidae, wolves (object), 
% Mosquite (event) and Wing (object), 
% Clock (event) and Analog clock (object) 
(Emergency vehicle (\textit{sound}); Ambulance (\textit{object})), 
(Grey wolf; Canidae, wolves), 
(Mosquite; Wing), and 
(Clock; Analog clock) are effectively individually clustered.
In addition, similar audio events and visual objects are located closely, like (Male singing; CD players, Speakers), (Mechanical fan, Squeak; Recreational vehicle, Tram).
The efficient aggregation of various event-object pairs in Fig.~\ref{w_s_e} illustrates that contrastive learning-based CEOA aligns fine-grained information of audio events and visual objects across modalities.

% \vspace{-0.1cm}
\section{CONCLUSION}\label{s4}

% \vspace{-0.2cm}
To exploit the fine-grained information of audio-visual events-object in diverse real-life scenes and to coordinate the implicit relationship between audio events and visual objects, 
this paper proposes a multi-branch AVSC model equipped with CEOA for event-object alignment and SF for cross-modal fusion.
Experiments show that 1) the proposed contrastive learning-based CEOA aligns fine-grained information of audio events and visual objects, and SF successfully fuses cross-modal information;
2) The influence of contrastive learning in model training is beneficial to the model’s recognition of fine-grained event-object pairs in audio-visual scenes; 
3) Even without using additional datasets and data augmentation tricks, the proposed model shows competitive performance.

% \vspace{-0.1cm}
\label{sec:refs}
\bibliographystyle{IEEEbib}
\bibliography{mybib}

\end{document}